\documentstyle[twoside]{article}

\catcode`\@=11
\long\def\@makefntext#1{
\protect\noindent \hbox to 3.2pt {\hskip-.9pt  
$^{{\eightrm\@thefnmark}}$\hfil}#1\hfill}		

\def\@makefnmark{\hbox to 0pt{$^{\@thefnmark}$\hss}}	
	
\def\ps@myheadings{\let\@mkboth\@gobbletwo
\def\@oddhead{\hbox{}
\rightmark\hfil\eightrm\thepage}   
\def\@oddfoot{}\def\@evenhead{\eightrm\thepage\hfil
\leftmark\hbox{}}\def\@evenfoot{}
\def\sectionmark##1{}\def\subsectionmark##1{}}

\oddsidemargin=\evensidemargin
\newcounter{sectionc}\newcounter{subsectionc}\newcounter{subsubsectionc}
\renewcommand{\section}[1] {\vspace{12pt}\addtocounter{sectionc}{1} 
\setcounter{subsectionc}{0}\setcounter{subsubsectionc}{0}\noindent 
	{\tenbf\thesectionc. #1}\par\vspace{5pt}}
\renewcommand{\subsection}[1] {\vspace{12pt}\addtocounter{subsectionc}{1} 
	\setcounter{subsubsectionc}{0}\noindent 
	{\bf\thesectionc.\thesubsectionc. {\kern1pt \bfit #1}}\par\vspace{5pt}}
\renewcommand{\subsubsection}[1] {\vspace{12pt}\addtocounter{subsubsectionc}{1}
	\noindent{\tenrm\thesectionc.\thesubsectionc.\thesubsubsectionc.
	{\kern1pt \tenit #1}}\par\vspace{5pt}}

\newcounter{appendixc}
\newcounter{subappendixc}[appendixc]
\newcounter{subsubappendixc}[subappendixc]
\renewcommand{\thesubappendixc}{\Alph{appendixc}.\arabic{subappendixc}}
\renewcommand{\thesubsubappendixc}
	{\Alph{appendixc}.\arabic{subappendixc}.\arabic{subsubappendixc}}

\renewcommand{\appendix}[1] {\vspace{12pt}
        \refstepcounter{appendixc}
        \setcounter{figure}{0}
        \setcounter{table}{0}
        \setcounter{lemma}{0}
        \setcounter{theorem}{0}
        \setcounter{corollary}{0}
        \setcounter{definition}{0}
        \setcounter{equation}{0}
        \renewcommand{\thefigure}{\Alph{appendixc}.\arabic{figure}}
        \renewcommand{\thetable}{\Alph{appendixc}.\arabic{table}}
        \renewcommand{\theappendixc}{\Alph{appendixc}}
        \renewcommand{\thelemma}{\Alph{appendixc}.\arabic{lemma}}
        \renewcommand{\thetheorem}{\Alph{appendixc}.\arabic{theorem}}
        \renewcommand{\thedefinition}{\Alph{appendixc}.\arabic{definition}}
        \renewcommand{\thecorollary}{\Alph{appendixc}.\arabic{corollary}}
        \renewcommand{\theequation}{\Alph{appendixc}.\arabic{equation}}
        \noindent{\tenbf Appendix \theappendixc #1}\par\vspace{5pt}}
\newcommand{\subappendix}[1] {\vspace{12pt}
        \refstepcounter{subappendixc}
        \noindent{\bf Appendix \thesubappendixc. {\kern1pt \bfit #1}}
	\par\vspace{5pt}}
\newcommand{\subsubappendix}[1] {\vspace{12pt}
        \refstepcounter{subsubappendixc}
        \noindent{\rm Appendix \thesubsubappendixc. {\kern1pt \tenit #1}}
	\par\vspace{5pt}}

\newcommand{\textlineskip}{\baselineskip=13pt}
\newcommand{\smalllineskip}{\baselineskip=10pt}

\def\eightcirc{
\begin{picture}(0,0)
\put(4.4,1.8){\circle{6.5}}
\end{picture}}
\def\eightcopyright{\eightcirc\kern2.7pt\hbox{\eightrm c}} 

\newcommand{\copyrightheading}[1]
	{\vspace*{-2.5cm}\smalllineskip{\flushleft
	{\footnotesize International Journal of Modern Physics A, #1}\\
	{\footnotesize $\eightcopyright$\, World Scientific Publishing
	 Company}\\
	 }}

\def\abstracts#1#2#3{{
	\centering{\begin{minipage}{4.5in}\baselineskip=10pt\footnotesize
	\parindent=0pt #1\par 
	\parindent=15pt #2\par
	\parindent=15pt #3
	\end{minipage}}\par}} 

\renewenvironment{thebibliography}[1]
	{\frenchspacing
	 \ninerm\baselineskip=11pt
	 \begin{list}{\arabic{enumi}.}
	{\usecounter{enumi}\setlength{\parsep}{0pt}
	 \setlength{\leftmargin 12.7pt}{\rightmargin 0pt} 
	 \setlength{\itemsep}{0pt} \settowidth
	{\labelwidth}{#1.}\sloppy}}{\end{list}}

\newcounter{itemlistc}
\newcounter{romanlistc}
\newcounter{alphlistc}
\newcounter{arabiclistc}

\newcommand{\fcaption}[1]{
        \refstepcounter{figure}
        \setbox\@tempboxa = \hbox{\footnotesize Fig.~\thefigure. #1}
        \ifdim \wd\@tempboxa > 5in
           {\begin{center}
        \parbox{5in}{\footnotesize\smalllineskip Fig.~\thefigure. #1}
            \end{center}}
        \else
             {\begin{center}
             {\footnotesize Fig.~\thefigure. #1}
              \end{center}}
        \fi}

\newcommand{\tcaption}[1]{
        \refstepcounter{table}
        \setbox\@tempboxa = \hbox{\footnotesize Table~\thetable. #1}
        \ifdim \wd\@tempboxa > 5in
           {\begin{center}
        \parbox{5in}{\footnotesize\smalllineskip Table~\thetable. #1}
            \end{center}}
        \else
             {\begin{center}
             {\footnotesize Table~\thetable. #1}
              \end{center}}
        \fi}

\def\@citex[#1]#2{\if@filesw\immediate\write\@auxout
	{\string\citation{#2}}\fi
\def\@citea{}\@cite{\@for\@citeb:=#2\do
	{\@citea\def\@citea{,}\@ifundefined
	{b@\@citeb}{{\bf ?}\@warning
	{Citation `\@citeb' on page \thepage \space undefined}}
	{\csname b@\@citeb\endcsname}}}{#1}}

\newif\if@cghi
\def\cite{\@cghitrue\@ifnextchar [{\@tempswatrue
	\@citex}{\@tempswafalse\@citex[]}}
\def\citelow{\@cghifalse\@ifnextchar [{\@tempswatrue
	\@citex}{\@tempswafalse\@citex[]}}
\def\@cite#1#2{{$\null^{#1}$\if@tempswa\typeout
	{IJCGA warning: optional citation argument 
	ignored: `#2'} \fi}}

\def\pmb#1{\setbox0=\hbox{#1}
	\kern-.025em\copy0\kern-\wd0
	\kern.05em\copy0\kern-\wd0
	\kern-.025em\raise.0433em\box0}

\def\fnt#1#2{\footnotetext{\kern-.3em
	{$^{\mbox{\scriptsize #1}}$}{#2}}}

\def\fpage#1{\begingroup
\voffset=.3in
\thispagestyle{empty}\begin{table}[b]\centerline{\footnotesize #1}
	\end{table}\endgroup}

\def\runninghead#1#2{\pagestyle{myheadings}
\markboth{{\protect\footnotesize\it{\quad #1}}\hfill}
{\hfill{\protect\footnotesize\it{#2\quad}}}}
\font\tenrm=cmr10
\font\tenit=cmti10 
\font\tenbf=cmbx10
\font\bfit=cmbxti10 at 10pt
\font\ninerm=cmr9

\font\eightrm=cmr8

\def\qed{\hbox{${\vcenter{\vbox{			
   \hrule height 0.4pt\hbox{\vrule width 0.4pt height 6pt
   \kern5pt\vrule width 0.4pt}\hrule height 0.4pt}}}$}}

\begin{document}

\runninghead
{Quintessence Model and Cosmic Microwave Background}
{Quintessence Model and Cosmic Microwave Background}

\normalsize\textlineskip
\thispagestyle{empty}
\setcounter{page}{1}

\copyrightheading{}			

\vspace*{0.88truein}

\fpage{1}
\centerline{\bf QUINTESSENCE MODEL AND COSMIC}
\vspace*{0.035truein}
\centerline{\bf MICROWAVE BACKGROUND}
\vspace*{0.37truein}
\centerline{\footnotesize PAUL H. FRAMPTON } 
\vspace*{0.015truein}
\centerline{\footnotesize\it Institute of Field Physics, 
Department of Physics and Astronomy,}
\vspace*{0.015truein}
\centerline{\footnotesize\it University of North Carolina,
Chapel Hill, NC 27599-3255}
\vspace*{0.225truein}

\vspace*{0.21truein}
\abstracts{
A particular kind of quintessence is considered, with
equation of motion $p_Q/\rho_Q = -1$, corresponding
to a cosmological
term with time-dependence $\Lambda(t) =
\Lambda(t_0) (R(t_0)/R(t))^{P}$ and we examine
how values of $\Omega_m$ and $\Omega_{\Lambda}$
depend on $P$.
}{}{}
\bigskip
\bigskip


\vspace*{1pt}\textlineskip	
\vspace*{-0.5pt}
\noindent

\bigskip

In this talk I summarize the paper\cite{CDFN}.
We shall investigate the position of the first Doppler peak in the Cosmic
Microwave Background (CMB) analysis using results published earlier\cite{FNR}. 

The combination of the information about the first Doppler peak and the complementary
analysis of the deceleration parameter derived from observations of the high-red-shift
supernovae leads to fairly precise values for the cosmic
parameters $\Omega_m$ and $\Omega_{\Lambda}$. We shall therefore also investigate the
effect of quintessence on the values of these parameters.

\bigskip

To introduce our quintessence model as a time-dependent cosmological term,
we start from the Einstein equation:

\begin{equation}
R_{\mu\nu} - \frac{1}{2} R g_{\mu\nu} = \Lambda(t) g_{\mu\nu} + 8 \pi G T_{\mu\nu} = 8 \pi G {\cal T}_{\mu\nu}
\label{einstein}
\end{equation}

\noindent where $\Lambda(t)$ depends on time as will be specified later
and $T_{\nu}^{\mu} = {\rm diag} (\rho, -p, -p, -p)$.
Using the Robertson-Walker metric, the `00' component of Eq.(\ref{einstein})
is

\begin{equation}
\left( \frac{\dot{R}}{R} \right)^2 + \frac{k}{R^2} = \frac{8 \pi G \rho}{3} +
 \frac{1}{3}\Lambda
\label{00}
\end{equation}

\noindent while the `ii' component is
\begin{equation}
2\frac{\ddot{R}}{R} + \frac{\dot{R}^2}{R^2} + \frac{k}{R^2} = -8 \pi G p + \Lambda
\label{ii}
\end{equation}
Energy-momentum conservation follows from Eqs.(\ref{00},\ref{ii}) because of the Bianchi identity
$D^{\mu} (R_{\mu\nu} - \frac{1}{2} g_{\mu\nu}) = D^{\mu} (\Lambda g_{\mu\nu} + 8\pi G T_{\mu\nu})
= D^{\mu} {\cal T}_{\mu\nu} = 0$.

\bigskip

Note that the separation of ${\cal T}_{\mu\nu}$ into two terms, one involving $\Lambda(t)$,
as in Eq(\ref{einstein}), is not meaningful except in a phenomenological sense because of energy conservation.

\bigskip

In the present cosmic era, denoted by the subscript `0', Eqs.(\ref{00},\ref{ii}) become respectively:

\begin{equation}
\frac{8\pi G}{3} \rho_0 = H_0^2 + \frac{k}{R_0^2} - \frac{1}{3} \Lambda_0
\label{00now}
\end{equation}
\begin{equation}
- 8 \pi G p_0 = - 2 q_0 H_0^2 + H_0^2 + \frac{k}{R_0^2} - \Lambda_0
\label{iinow}
\end{equation}
where we have used $q_0 = - \frac{\ddot{R}_0}{R_0 H_0^2}$ and $H_0 = \frac{\dot{R}_0}{R_0}$.

For the present era, $p_0 \ll \rho_0$ for cold matter and then Eq.(\ref{iinow}) becomes:

\begin{equation}
q_0 = \frac{1}{2} \Omega_{M} - \Omega_{\Lambda}
\label{decel}
\end{equation}
where $\Omega_{M} = \frac{8 \pi G \rho_0}{3 H_0^2}$ and $\Omega_{\Lambda} = \frac{\Lambda_0}{3 H_0^2}$.

\bigskip
\bigskip

Now we can introduce the form of $\Lambda(t)$ we shall assume by writing

\begin{equation}
\Lambda(t) = b R(t)^{-P}
\end{equation}
where $b$ {\bf is} a constant and the exponent $P$ we shall study for the
range $0 \leq P < 3$. This motivates the introduction of the new variables

\begin{equation}
\tilde{\Omega}_M = \Omega_M - \frac{P}{3 - P} \Omega_{\Lambda} , ~~~~\tilde{\Omega}_{\Lambda}
= \frac{3}{3 - P} \Omega_{\Lambda}
\label{tilde}
\end{equation}

\noindent It is unnecessary to redefine $\Omega_C$
because $\tilde{\Omega}_M + \tilde{\Omega}_{\Lambda}
= \Omega_M + \Omega_{\Lambda}$. 

\bigskip

The equation for the first Doppler peak incorporating the possibility of non-zero $P$
is found to be the following modification 
for $\Omega_C=0$

\bigskip

\begin{equation}
l_1 = \pi \left( \frac{R_t}{R_0} \right)
\left[\tilde{\Omega}_M  \left( \frac{R_0}{R_t} \right)^3 + \tilde{\Omega}_{\Lambda}
\left( \frac{R_0}{R_t} \right)^P
 \right]^{1/2}
\int_1^{\frac{R_0}{R_t}} \frac{dw}{\sqrt{\tilde{\Omega}_M w^3 + \tilde{\Omega}_{\Lambda} w^P}}
\label{l1Pflat}
\end{equation}

\bigskip

\noindent The dependence of $l_1$ on $P$ is illustrated 
graphically by figures in \cite{CDFN}. Further discussion of the dependence of $l_1$
on $\Omega_C$, using the formulas of \cite{FNR}, is
given in \cite{weinberg}.

\bigskip

\noindent We have introduced $P$ as a parameter which is real and with
$0 \leq P < 3$.
For $P \rightarrow 0$ we regain the standard cosmological model. But now
we must investigate other restrictions already necessary for $P$ before precision
cosmological measurements restrict its range even further.

\bigskip

Only for certain $P$ is it possible to extrapolate the cosmology
consistently for all $0 < w = (R_0/R) < \infty$. For example, in the flat case $\Omega_C = 0$
which our universe seems to approximate, the formula for the expansion rate is

\begin{equation}
\frac{1}{H_0^2} \left( \frac{\dot{R}}{R} \right)^2 = \tilde{\Omega}_M w^3 + \tilde{\Omega}_{\Lambda} w^P
\label{flatexp}
\end{equation}

\noindent This is consistent as a cosmology only if the right-hand side has no zero for a real
positive $w = \hat{w}$. The root $\hat{w}$ is

\begin{equation}
\hat{w} = \left( \frac{ 3(1 - \Omega_M)}{P - 3 \Omega_M} \right)^{\frac{1}{3 - P}}
\label{wcrit}
\end{equation}

\bigskip

\noindent If $0 < \Omega_M < 1$, consistency requires that $P < 3 \Omega_M$.

\bigskip

Another constraint on the cosmological model is provided by nucleosynthesis which requires that
the rate of expansion for very large $w$ does not differ too much from that
of the standard model.

\bigskip

The expansion rate for $P = 0$ coincides for large $w$ with that of the standard model so it
is sufficient to study the ratio:

\begin{eqnarray}
(\dot{R}/R)_P^2/(\dot{R}/R)_{P=0}^2
&
\stackrel{w \rightarrow \infty}{\rightarrow}
& (3 \Omega_{M} - P)/((3 - P) \Omega_{M})\\
& \stackrel{w \rightarrow \infty}{\rightarrow}& (4 \Omega_{R} - P)/((4 - P)\Omega_{R})
\end{eqnarray}

\noindent where the first limit is for matter-domination and the second is for radiation-domination
(the subscript R refers to radiation).

\bigskip

\noindent The constraints of avoiding a bounce ($\dot{R} = 0$) in the past, and then requiring
consistency with BBN leads to $0 < P < 0.2$.

\bigskip
\noindent Clearly, from the point of view of inflationary cosmology, the precise
vanishing of $\Omega_C = 0$ is a crucial test and its confirmation
will be facilitated by comparison models such as the present one.


\begin{thebibliography}{99}
\bibitem{CDFN}
J.L. Crooks, J.O. Dunn, P.H. Frampton and Y.J. Ng. {\tt astro-ph/0005406}
\bibitem{FNR}
P.H. Frampton, Y.J. Ng and R.M Rohm,
Mod. Phys. Lett {\bf A13,} 2541 (1998).
{\tt astro-ph/9806118}.
\bibitem{weinberg}
S. Weinberg, {\tt astro-ph/0006276}.
\end{thebibliography}
\end{document}